\documentclass[showpacs,floatfix,aps,pre,twocolumn]{revtex4-1}
\usepackage[utf8]{inputenc}
\usepackage{amsmath}
\usepackage{amssymb}
\usepackage{refcount}
\usepackage{siunitx}
\usepackage{graphicx}
\usepackage[english]{babel}
\usepackage{bm}
\usepackage{color}
\usepackage[capitalise]{cleveref}
\usepackage{etoolbox}
\usepackage{appendix}
\makeatletter
\appto{\appendix}{%
	\@ifstar{\def\theequation@prefix{A.}}%
	{}%
}
\makeatother

\newcommand{\bx}{\mathbf{x}}
\newcommand{\bp}{\mathbf{p}}
\newcommand{\bs}{\mathbf{s}}
\newcommand{\bE}{\mathbf{E}}
\newcommand{\bB}{\mathbf{B}}
\newcommand{\bsigma}{\bm{\sigma}}

\newcommand{\bpi}{\bm{\pi}}

\newcommand{\hE}{\hat{E}}
\newcommand{\hpi}{\hat{\pi}}
\newcommand{\hB}{\hat{B}}
\newcommand{\hT}{\hat{T}}
\newcommand{\hp}{\hat{p}}
\newcommand{\hx}{\hat{x}}
\newcommand{\hO}{\hat{O}}

\newcommand{\tB}{\tilde{\bB}}
\newcommand{\tE}{\tilde{\bE}}

\newcommand{\ioverhbar}{\frac{i}{\hbar}}
\newcommand{\textinttau}{\textstyle \int_0^1 d\tau \, }

\newcommand{\xpleftright}{\overset{\leftarrow}{\partial_x} \cdot \overset{\rightarrow}{\partial_p}}

\newcommand{\pd}[2]{\frac{\partial #1}{\partial #2}}

\newcommand{\pres}{\ensuremath{p_{\textnormal{res}}}}
\newcommand{\pth}{\ensuremath{p_{\textnormal{th}}}}

\DeclareMathOperator{\Tr}{Tr}
\renewcommand{\Im}{\operatorname{Im}}
\renewcommand{\Re}{\operatorname{Re}}

\begin{document}
	
\title{Short-scale quantum kinetic theory including spin-orbit interactions} 
\author{R. Ekman, H. Al-Naseri, J. Zamanian and G. Brodin}
\affiliation{Department of Physics, Ume{\aa } University, SE--901 87 Ume{\aa}, Sweden}

\begin{abstract}
	We present a quantum kinetic theory for spin-$1/2$ particles, including the spin-orbit interaction, retaining  particle dispersive effects to all orders in $\hbar$, based on a gauge-invariant Wigner transformation.
	Compared to previous works, the spin-orbit interaction leads to a new term in the kinetic equation, containing both the electric and magnetic fields.
	Like other models with spin-orbit interactions, our model features ``hidden momentum''. 
	As an example application, we calculate the dispersion relation for linear electrostatic waves in a magnetized plasma, and electromagnetic waves in a unmagnetized plasma.
    In the former case, we compare the Landau damping due to spin-orbit interactions to that due to the free current.
    We also discuss our model in relation to previously published works.
\end{abstract}

\maketitle

\section{Introduction}

Dense plasmas, where quantum effects are important, can, for example, be found in different types of solid-state plasmas~\cite{Shukla-Eliasson-RMP,manfredi2015solid}, various astrophysical environments~\cite{chabrier2002dense,uzdensky2014extreme}, and certain forms of laser-plasma interactions~\cite{DiPiazza2012,weber2018p3}.
Such plasmas have been modelled using hydrodynamic~\cite{Haas-book} and kinetic equations~\cite{bonitz2016quantum}, where, in the present paper, the focus is on the latter class of theories.
Quantum kinetic equations can be derived, for example, from the Green's function formalism~\cite{kadanoff1962quantum,keldysh1965diagram} or from the density matrix approach~\cite{hohenester1997density,PhysRevE.96.023207}.
Recently, several kinetic models for plasmas have been put forward, within
the Hartree approximation~\cite{ZamanianNJP,PhysRevE.96.023207,andreev2016soliton,hurst2014semiclassical}
or the Hartree-Fock approximation~\cite{ekman2015exchange,andreev2016landau,haas2019exchange}.
Depending on the scope of the theory, the density matrix can be based on the Schr\"{o}dinger equation~\cite{bonitz2016quantum}, the Pauli equation~\cite{ZamanianNJP}, or the Dirac equation, where the latter can be applied in the weakly~\cite{asenjo2012semi} or fully~\cite{PhysRevE.96.023207} relativistic approximation.

In order to produce a quantum kinetic theory of electrons from the Dirac equation, certain restrictions need to be applied.
Firstly, pair-production must be negligible, such that a Foldy-Wouthuysen transformation~\cite{foldy1950dirac,Silenko2008} can be applied to separate electron states from positron states.
This puts restrictions on the maximal electric field strength, which should be well below the critical field $E_\text{cr} = m^2 c^3/|q|\hbar $, and the characteristic spatial scales of the fields, which should be much longer than the Compton length $ L_c = \hbar/m c$.
Here $m$ and $q$ are the electron mass and charge, respectively, $c$ is the speed of light in vacuum, and $\hbar $ is the reduced Planck's constant.
Given these restrictions, there are two different regimes that can be studied.
Firstly, there is the fully relativistic regime, where the relativistic factor $\gamma $ fulfills $\gamma -1\sim 1$, in which case the de Broglie length $\lambda_\text{dB} = \hbar /p_{\mathrm{ch}}$ is of the same order as $L_{c}$.
Here $p_{\mathrm{ch}}$ is the characteristic momentum of the electrons -- determined by the Fermi temperature $T_{F}$ or the thermodynamic temperature $T$ whichever is larger.
Since, according to the applicability conditions, this means that the spatial scales must be longer than the de Broglie length, particle dispersive effects cannot be included in the fully relativistic regime~\cite{PhysRevE.96.023207}.
However, for the second case, the weakly relativistic regime with $\gamma -1\ll {1}$, we have $\lambda _{dB}\gg L_{c}$, and
as a consequence, a theory including particle dispersive effects based on the Dirac equation can be formulated.
This is the main goal of the present paper.
We thus generalize the work presented in Ref.~\cite{ZamanianNJP} based on the Pauli equation, by including several new effects, such as spin-orbit interaction, Thomas precession and the polarization currents associated with the spin.
We also generalize the results of Ref.~\cite{asenjo2012semi}, by also covering the short scale physics down to spatial scales of the order of the de Broglie length.

Our approach is as follows.
We start from the Pauli Hamiltonian including the spin-orbit interaction.
In \cref{sec:calculations}, we apply a combined Wigner- and  Q-transform (for the spin) using the gauge-independent approach of Stratonovich~\cite{stratonovich1956gauge,serimaa}, to formulate our scalar kinetic theory.

In \cref{sec:dispersion}, in order to demonstrate the usefulness of the present theory, we have calculated two examples from linearized theory: electromagnetic waves in a non-magnetized plasma and electrostatic waves propagating parallel to an external magnetic field.
Classically, an external magnetic field does not affect parallel propagating electrostatic waves.
By contrast, the present model predicts a condition, depending on the magnetic field, for resonant wave-particle interaction, generalizing those of previous works.
We use this result to calculate the magnetic field dependence of the damping rate.

Finally, in \cref{sec:discussion}, we compare our model with some theories from the recent research literature.

\section{The gauge invariant Wigner function and the spin transformation}
\label{sec:calculations}

We will consider the semi-relativistic Pauli Hamiltonian \begin{equation}
	\hat{H} = \frac{\hat{\bpi}^2}{2m} + q\hat{\phi} - \mu_e \bsigma \cdot \hat{\bB} + \frac{\mu_e}{2mc^2}\left( \hat{\bpi} \times \hat{\bE} - \hat{\bE} \times \hat{\bpi} \right)\cdot \bsigma
    \label{eq:hamiltonian}
\end{equation}
including the spin-orbit interaction.
Here $\hat{\bpi} = \hat{\bp} - q\hat{\mathbf A}$ is the gauge invariant momentum operator, $\hat{\phi}, \hat{\mathbf A}$ are the scalar and vector potentials and $\mu_e$ is the electron magnetic moment.
Note that the spin-orbit interaction is written in symmetric form so as to make the Hamiltonian Hermitian -- which is necessary to have unitary time evolution.

Our goal is to formulate a scalar quantum kinetic theory based on the gauge invariant Wigner function~\cite{stratonovich1956gauge,serimaa}, that is, \begin{equation}
	W(\bx, \bp, t)_{\alpha\beta}
    = \operatorname{tr} \left[ \hat{W}(\bx, \bp, t) \rho_{\alpha\beta}\right]
    = \int d^3 \mathbf{r} \, \langle \mathbf{r} | \hat{W} \hat{\rho}_{\alpha\beta} | \mathbf{r} \rangle,
    \label{eq:wigner}
\end{equation}
where $\alpha,\beta$ are the Pauli spin indices, and this trace is only over the spatial degrees of freedom.
We work in the Heisenberg picture so that the operator $\hat{W}(\bx, \bp, t)$ is time-dependent, but the density matrix $\rho$ is not.
The operator $\hat{W}(\bp, \bx, t)$ can be expressed as a Fourier transform
\begin{align}
    \hat{W}(\bx, \bp, t)
    & = \left(\mathcal{F}(\hT)\right)(\bx, \bp) \notag \\
    & = \int \frac{d^3 u}{(2\pi\hbar)^3}\frac{d^3 v}{(2\pi\hbar)^3}
        e^{-\ioverhbar (\mathbf{u} \cdot \bp + \mathbf{v} \cdot \bx)}
        \,
        \hT(\mathbf u, \mathbf v),
\end{align}
where the operator $\hat{T}(\mathbf u, \mathbf v, t)$ is given by
\begin{subequations}
	\begin{equation}
		\hT(\mathbf u, \mathbf v, t) = \exp\left[ \ioverhbar (\mathbf u \cdot \hat{\bpi} + \mathbf v \cdot \hat{\bx}) \right].
	\end{equation}
    \label{eq:T-start}
\end{subequations}

To obtain a scalar function from the matrix-valued $\hat{W}_{\alpha\beta}$, one can use the spin transform
\begin{align}
	f(\bx, \bp, \mathbf{s})
    & = \frac{1}{4\pi} \operatorname{tr}[ (1 + \mathbf{s}\cdot \bsigma) W ] \notag\\
    & = \frac{1}{4\pi} \sum_{\alpha,\beta} [\delta_{\alpha\beta} + \bs \cdot \bsigma_{\alpha\beta}] W_{\beta\alpha}
    \label{eq:spin-transform}
\end{align}
which is a Husimi $Q$-function for the spin~\cite{husimi1940some}.
This transform was used in Ref.~\cite{ZamanianNJP} for a theory including only the magnetic dipole interaction.

The remainder of this section presents the calculations involved in deriving an evolution equation for $f$, with the conclusion presented in \cref{sec:calculations-conc}.
The calculations in this section are similar to those in Ref.~\cite{serimaa}, where the spin-independent terms in the kinetic equation are found.

Our notation is as follows.
Throughout, quantities with hats are operators and quantities without hats are $c$-numbers, except the Pauli spin operator $\bsigma$ which is always an operator (a finite dimensional one, i.e., a matrix).
We will also use the summation convention for repeated indices, and use commas to denote derivatives, e.g., $E_{i,j} = \partial E_i / \partial x_j$.

Various other forms of the operator $\hT$ (Baker-Campbell-Haussdorff formulas) can be found~\cite{serimaa} and will be useful in the following:
\begin{widetext}
    \setcounterref{equation}{eq:T-start}
    \addtocounter{equation}{-1}
    \begin{subequations}
        \stepcounter{equation}
	\begin{align}
		\hT(u,v) & =
		\exp\left[\frac{i}{2\hbar} u\cdot v   \right]
		\exp\left[\
		\ioverhbar v\cdot \big(\hx - q u \cdot \textinttau \hat{A}(\hx + \tau u)\big)
		\right]
		\exp\left[\ioverhbar u\cdot \hat p\right] \label{bch:x-p} \\
		& =
		\exp\left[-\frac{i}{2\hbar} u\cdot v   \right]
		\exp\left[\ioverhbar u\cdot \hat p\right] \label{bch:p-x}
		\exp\left[\
		\ioverhbar v\cdot \big(\hx - q u \cdot \textinttau \hat{A}(\hx + \tau u)\big)
		\right] \\
		& = \exp\left[\frac{i}{2\hbar} u\cdot v   \right]
		\exp\left[\
		\ioverhbar v\cdot \hx
		\right]
		\exp\left[\ioverhbar u\cdot \hat \pi\right] \label{bch:x-pi} \\
		& = \exp\left[-\frac{i}{2\hbar} u\cdot v   \right]
		\exp\left[\
		\ioverhbar u\cdot \hat{\pi}
		\right]
		\exp\left[\ioverhbar v\cdot \hat x\right] \label{bch:pi-x}
        .
	\end{align}
\end{subequations}
    \setcounterref{equation}{eq:spin-transform}
\end{widetext}
One can note that these occur in pairs related by operator orderings.

The time evolution of $\hat{W}$ is given by the Heisenberg equation of motion, \begin{equation}
	\frac{d}{dt}\hat{W} = \pd{}{t} \hat{W} - \ioverhbar [\hat{W}, \hat{H}].
\end{equation}
Here the only effect of the partial time derivative is to give the $\partial_t \mathbf A$ contribution to the electric field in the Lorentz force.
In the Heisenberg picture, the operator $\bsigma$ is time-dependent, so we must remember to include $d\bsigma/dt$ when deriving the evolution equation for $f$.
Hence, if we let $\hat{S} = \frac{1}{4\pi} (1 + \mathbf s \cdot \bsigma)$, then \begin{equation}
	f = \Tr [ \hat{S} \hat{W} \rho ]
\end{equation}
i.e. $f$ is the expectation value of $\hat{S} \hat{W}$.
Note that $\hat{S}$ and $\hat{W}$ are both Hermitian and commute, which implies that $f$ is real.
The Heisenberg equation of motion then gives \begin{equation}
	\partial_t f = \Tr \left[ \left( \hat{S} \partial_t \hat{W} - \ioverhbar \hat{S} [\hat{W}, \hat{H}] - \ioverhbar [\hat{S}, \hat{H}] \hat{W}   \right) \rho \right]
\end{equation}
where in this and the previous equation, the trace is over both the spin indices and the spatial degrees of freedom.

The commutator is linear in $\hat{H}$, and the terms in the evolution equation for $\hat{W}$ corresponding to the lowest order (``non-relativistic'') part of the Hamiltonian have already been computed in~\cite{ZamanianNJP}.
Therefore we will here be concerned only with the spin-orbit contribution 
\begin{equation}
	\hat{H}_{\textrm{SO}} = 
	\frac{\mu_e	}{2 m c^2}
	(\hat{\bpi} \times \hat{\bE} - \hat{\bE} \times \hat{\bpi} ) \cdot \bsigma	.
	\label{eqhso}
\end{equation}
We will the denote the contribution from this Hamiltonian by $ \left( \partial_t f \right)_{\textrm{SO}}  $.

\newcommand{\SO}{\ensuremath{\textnormal{SO}}}

The spin-orbit interaction Hamiltonian has the form $\bsigma \cdot \hat{\mathbf V}$, and hence we find
\begin{align}
	[\hat{W}, \hat{H}_\SO] & = \sigma_i [\hat{W} , \hat{V}_i] \\
	[\hat{S}, \hat{H}_\SO] & = 2i \, s_i \epsilon_{ijk} \hat{V}_j \sigma_k.
\end{align}
Then using $\sigma_i\sigma_j = \delta_{ij} + i \epsilon_{ijk}$, we get \begin{equation}
	\hat{S} [\hat{W}, \hat{H}_{\textrm{SO}}] =
	(\sigma_j + s_j  + i\epsilon_{ijk} s_i \sigma_k) [\hat{W} , \hat{V}_j].
    \label{eq:STcomH}
\end{equation}
For the second commutator, we write $\hat{\mathbf V}\hat{W}$ as the sum of its symmetric and anti-symmetric parts
\begin{multline}
	2i\epsilon_{ijk} s_i \hat{V}_j \sigma_k \hat{W}
    \\
    = i\epsilon_{ijk} s_i \sigma_k \left(
        \big(\hat{V}_j\hat{W} + \hat{W}\hat{V}_j \big)
        +  [\hat{V_j}, \hat{W}]
    \right)
\end{multline}
and clearly, the second part cancels with cross product part of $\hat{S} [\hat{W}, \hat{H}_{\textrm{SO}}]$ in \cref{eq:STcomH}.

Now we just need to use
\begin{equation}
	\partial_{s_i} \hat{S} = \sigma_i - s_i s_j \sigma_j , 
\end{equation}
which implies
\begin{align}
	(\partial_{s_i} + s_i) \hat{S} & = \sigma_i + s_i \\
	\epsilon_{ijk} \partial_{s_i} s_j \hat{S} & = \epsilon_{ijk} \sigma_i s_j,
\end{align}
and thus
    \begin{multline}
		\left( \partial_t f \right)_\textrm{SO} =
        - \ioverhbar \Tr \Big[ \big( 
        (\partial_{s_i} + s_i) \hat{S} [\hat{W}, \hat{V_i} ]
		\\
        + \epsilon_{ijk} \partial_{s_i} s_j \hat{S}[\hat{W}, \hat{V_k}]_+
    \big) \rho
    \Big] . 
    \end{multline}

The remaining, laborious, task is to evaluate and take the Fourier transform of $\hT (\hat{\bpi}\times \hat{\bE}) \pm (\hat{\bpi}\times \hat{\bE}) \hT $, then express the result in terms of differential operators -- functions of $\partial_p$ and $\partial_x$ -- acting on $\hat{W}$.
These operators can be taken outside the trace, so that they act on $f$, and this will give the kinetic equation.
Since
\begin{align}
	[\hat T, \hat{\bpi} \times \hat{\bE}] & = [\hat T, \hat{\bpi}] \times \hat{\bE} + \hat{\bpi} \times [\hat T, \hat{\bE}]
\end{align}
we need to evaluate the commutators with $\hat{\bpi}$ and $\hat{\bE}$.
This is sufficient, because for any operator $ \hat{O} $
\begin{equation}
	[\hat{W}, \hat{O}^\dagger] = - [\hat{W}, \hat{O}]^\dagger,
\end{equation}
since $\hat{W}$ is Hermitian,
and we get the commutator with the Hermitian ordering (which appears in \cref{eqhso}) by reading off the ``imaginary part'' of $[\hat{W}, \hat{\bpi}\times\hat{\bE}]$.

\subsection{The commutator $[\hat{W}, \hat{\bE}]$}
\label{ssub:sub1}

In \cref{ssub:sub1,ssub:sub2,ssub:sub3}, we will typically omit boldface for vector quantities; the type of each quantity should be apparent from context.

Using \cref{bch:x-p}, we can write the operator $\hat T$ with $\hat{\Theta}(u) = \exp(\ioverhbar u\cdot \hp)$ on the right.
Since $\hat p$ is the generator of translations, the operator $\Theta(u)$ acts like \begin{align}
    \hat{\Theta}(u) |x\rangle & = |x-u\rangle \\
    \hat{\Theta}(u) f(\hat x) & = f(\hx + u)\hat{\Theta}(u)
\end{align}
and thus, using \cref{bch:x-p} to put the translation operator on the left, with the other factors commuting with $\hE$,
    \begin{equation}
        [\hat T, \hat E_i]
        = \hT(u,v) \left( \hE_i(\hat x) - \hE_i(\hat x - u) \right)
        \label{eq:E-T-commutator}
    \end{equation}

Now we need an expression for the product $\hat{\bpi}\times [\hat T, \hat{\bE}]$ in terms of $\hat T$.
By putting the commutator in the form \cref{eq:E-T-commutator}, we can use any expression for $\hat T$ as is most appropriate.
The one that is most appropriate here is~\cref{bch:pi-x} since we can lift the result from Ref.~\cite{serimaa} their~(4.30b), and its Hermitian conjugate:
\begin{widetext}
    \begin{align}
        \hat{\pi}_i  \exp\left(\ioverhbar u\cdot \hat{\pi} \right) & =
        \left(
            \frac{\hbar}{i} \pd{}{u_i} - q \int_0^1 d\tau\,  (1-\tau) [u\times \hB(\hat{r} + \tau u)]_i
		\right) \exp\left(\ioverhbar u\cdot \hat{\pi} \right) \label{eq:pi-exp-pi} \\
        \exp\left(-\ioverhbar u\cdot \hat{\pi} \right)\hat{\pi}_i & =
        \exp\left(-\ioverhbar u\cdot \hat{\pi} \right)\left(
            -\frac{\hbar}{i} \overset{\leftarrow}{ \pd{}{u_i} } - q \int_0^1 d\tau\,  (1-\tau) [u\times \hB(\hat{r} + \tau u)]_i
        \right) .
    \end{align}

We thus have that the operator $\pi_i [\hat T, E_j]$ can be expressed as $\pi_i [\hat T, E_j]  = \hO^B_{ij} + \hO^D_{ij}$ where
{\allowdisplaybreaks
\begin{align}
    \hO^B_{ij} & = -q \int_0^1 d\tau\,
        (1-\tau) [u\times \hB(\hx + \tau u)]_i
        \hT(u,v) \left( \hE_j(\hat x) - \hE_j(\hat x - u) \right)
        \label{eq:O-B} \\
    \hO^D_{ij} & = e^{-\frac{i}{2\hbar} u\cdot v}\left[
        \frac{\hbar}{i} \pd{}{u_i} \exp\left(\ioverhbar u\cdot \hat \pi\right)
    \right]
        \exp\left( \ioverhbar v\cdot \hx \right)
        \left(\hE_j(\hx) - \hE_j(\hx - u)\right).
\end{align}
}

Now let $|x'\rangle, |x''\rangle$ be eigenstates of $\hat x$ (at the appropriate time).
Use~\cref{bch:x-p} so that the $\hx$ operators are on the left in $\hat T$.
\newcommand{\wilsonloop}{\ensuremath{
    \exp\left[\
        -\frac{iq}{\hbar} u \cdot \int_0^1 d\tau \,  \hat{A}(\hx + \tau u)
    \right]
}}
Let $\hat{Q}^B_{ij}$ be the Fourier transform in $u,v$ of $\hO^B_{ij}$. Then, letting $e^{i\hat{\Phi}} = \wilsonloop$ its matrix elements are
    \begin{multline}
        \langle x' | \hat{Q}^B_{ij} | x'' \rangle =
        \mathcal{F}\left[
        \langle x'| \hO^B_{ij} | x''\rangle \right] \\
        = -q\mathcal{F} \bigg[
            e^{\ioverhbar v\cdot (x' + u/2)}
            \int_0^1 d\tau \, (1-\tau) [u\times B(x' + \tau u)]_i  
            \langle x'|
            e^{i\Phi}
        e^{\ioverhbar u \cdot \hp} \left( \hE_j(\hat x) - \hE_j(\hat x - u) \right)
    |x''\rangle  \bigg]\\
    = -q\mathcal{F}\left[
     e^{\ioverhbar v\cdot (x' + u/2)}
    \int_0^1 d\tau \, (1-\tau) [u\times B(x' + \tau u)]_i 
    \langle x'|
    e^{i\Phi}
    \left( \hE_j(\hat x + u) - \hE_j(\hat x) \right)
|x''- u\rangle \right]\\
= -q\mathcal{F}\left[
    e^{\ioverhbar v\cdot (x' + u/2)}
    \int_0^1 d\tau \, (1-\tau) [u\times B(x' + \tau u)]_i 
    \big( E_j(x' + u) - E_j(x')\big) \langle x'|
    e^{i\Phi}
    |x''- u\rangle \right] .
    \end{multline}
\end{widetext}
At this point we see that the Fourier transform in $v$ can be carried out and will give an expression proportional to $\delta( x'-x + u/2 )$.
Hence the matrix elements of $\hat{Q}^B_{ij}(x,p)$ remain unchanged if we  change $x'$ to $x -u/2$ in the arguments of the fields.
In the integral over $\tau$, we also change variables to $\tau -1/2$ for a somewhat simpler expression.

We can then restore the operators making up $\hat T$, resulting in
\begin{widetext}
    \begin{multline}
        \langle x' | \hat{Q}^B_{ij} | x'' \rangle
        = -q\mathcal{F}
		\int_{-1/2}^{1/2} d\tau \, \left( \frac{1}{2}-\tau \right) [u\times B(x + \tau u)]_i 
        \big( E_j(x + u/2) - E_j(x - u/2)\big)
        \langle x'|
        e^{\frac{i}{2\hbar}u\cdot v}
        e^{\ioverhbar v \cdot x}
        e^{i\hat{\Phi}}
        |x''\rangle
        \\ =
        -q\mathcal{F} \left[
            \int_{-1/2}^{1/2} d\tau \, \left( \frac{1}{2}-\tau \right) [u\times B(x + \tau u)]_i 
            \big( E_j(x + u/2) - E_j(x - u/2)\big)
            \langle x' |
            \hT(u,v)
        |x'' \rangle \right]
    \end{multline}

Since the Fourier transform sends $u \mapsto i \hbar \partial_p$, we can now express this operator in terms of $\hat{W}$ and its derivatives.
First, use that
\begin{equation}
	 E_j(x + u/2) - E_j(x - u/2)
     = u_k \int_{-1/2}^{1/2} d\tau E_{j,k} (x + \tau u) , 
\end{equation}
and note that this quantity is real.
In the integral with $B$, $1/2$ is even and $-\tau$ is odd, so even and odd powers of $i\tau\hbar\partial_p$ in the expansion of $B$ survive the integration, respectively.
Since we want the imaginary part, it is the term with $-\tau$ that we should keep, and in conclusion
    \begin{equation}
        \operatorname{Im} \left( \hat{Q}^B_{ij}\right) =  -q\hbar^2
        \int_{-1/2}^{1/2} d\tau \, \tau  [\partial_p \times B(x + i\tau \hbar \partial_p )]_i
        \int_{-1/2}^{1/2}  d\sigma \, E_{j,k}(x + i\sigma\hbar \partial_p) \partial_{p_k} \hat{W}.
    \end{equation}

We now turn to $\hat{O}^D_{ij}$.
Note that from the expression for $\hT$ in \cref{bch:pi-x} we have \begin{equation}
	\pd{}{u_i} \hT =
		-\frac{i}{2\hbar} v_i \hT
		+ e^{-\frac{i}{2\hbar} u\cdot v} \pd{}{u_i}
        e^{\ioverhbar u\cdot \hat \pi} e^{\ioverhbar v\cdot \hat{x} }
\end{equation}
and hence
\begin{equation}
	\hat{O}^D_{ij} =
		 \hT\left[
		 	\frac{v_i}{2}
			 + \frac{\hbar}{i} \overset{\leftarrow}{\pd{}{u_i}}
		 	\right]
		\big(\hat{E}_j(\hat{x}) - \hat{E}_j(\hat{x} - u)\big).
\end{equation}
The first term can be handled using the argument above, except that the Fourier transform in $v$ will send $v_i$ to $i\hbar \partial_{x_i} \delta(x) $.
The result is that
    \begin{align}
        \mathcal{F} \left[\langle x' |
            \hT \frac{v_i}{2}   \big(\hat{E}_j(\hat{x}) - \hat{E}_j(\hat{x} - u)\big)
        |x'' \rangle \right]
        = -\frac{\hbar}{2i} \partial_{x_i} \left[
            \big( E_j(x + i \hbar \partial_p/2) - E_j(x - i\hbar \partial_p/2) \big)
        \langle x' | \hat{W} |x'' \rangle \right]
        \label{matels:v_i-term}
    \end{align}
but this quantity is real, and so makes no contribution to the evolution equation.
For the term with the $u$-derivative, we integrate by parts in the Fourier transform, i.e.,
\begin{equation}
    \mathcal{F}\left[
        \left(\frac{\partial}{\partial u_i} \hat{T} \right)
		 \big(\hat{E}_j(\hat{x}) - \hat{E}_j(\hat{x} - u)\big)
     \right]
	=
    \mathcal{F} \left[
        p_i \hT
        \big(\hat{E}_j(\hat{x}) - \hat{E}_j(\hat{x} - u)\big)
        - \frac{\hbar}{i}\hat{T} \hat{E}_{j,i}(\hat{x} - u)
		 \right] . 
	\label{eq:dT-du-integral}
\end{equation}
In the first term, $p_i$ can be taken outside the Fourier transform, and the remaining operator is treated like above resulting in
    \begin{equation}
        \mathcal{F}\left[ \langle x' | \hat{O}_{ij}^D | x'' \rangle \right] 
        = \left[ p_i \big(E_j(x + i\hbar \partial_p/2)  -  E_j(x - i\hbar\partial_p/2) \big) - \frac{\hbar}{i}E_{j,i}(x-i\hbar\partial_p/2)
        \right] \langle x' | \hat{W} |x''\rangle .
        \label{matels:OD}
    \end{equation}
We can write this in a somewhat simpler form as
\begin{equation}
	E_j(x + i\hbar \partial_p/2)  -  E_j(x - i\hbar\partial_p/2) = i\hbar (\partial_{x_k} \tilde{E}_j) \partial_{p_k}
    \label{eq:E-imag-part}
\end{equation}
where we have introduced the notation
\begin{equation}
	\tilde{E}_j = \int_{-1/2}^{1/2} E_j(x + i\hbar \tau \partial_p) \, d\tau.
    \label{eq:E-tilde-def}
\end{equation}
To summarize, we obtain
\begin{equation}
\operatorname{Im}(\hat{\bpi} \times [\hat{W}, \hat{\bE} ]) =
		 (\bp + \Delta\tilde{\bp}) \times \tilde{\bE}
	( \hbar\xpleftright )
	\hat{W}
    \label{eq:spin-orbit-force}
\end{equation}
where
\begin{equation}
	\Delta \tilde{\bp} = -\frac{q\hbar}{i}\partial_p \times \int_{-1/2}^{1/2} \, d\tau
	\tau \bB(\bx + i\hbar\tau\partial_p).
\end{equation}
The quantities $\tilde{\bE}$ and $\Delta\tilde{\bp}$ are the same as in Refs.~\cite{serimaa,ZamanianNJP}.
Since \cref{eq:spin-orbit-force} is to be contracted with the Pauli matrices, in the kinetic equation it represents the spin-orbit force.

\subsection{The commutator $[\hat{W}, \hat{\bpi}]$}
\label{ssub:sub2}
We use~\cref{bch:pi-x} for $\hat{T}$, so that
	\begin{align}
		[\hT, \hat{\pi}_i] E_j
        & = [-\hat{\pi}_i, \hT] E_j
          = -e^{-\frac{i}{2\hbar} u\cdot v} \left[
			\hat{\pi_i}, \exp(\ioverhbar u\cdot \hat{\pi}) \exp(\ioverhbar v\cdot \hat{x} )
		\right] E_j \notag \\
        & =  e^{-\frac{i}{2\hbar} u\cdot v}\left(
			-v_i + q \int_0^1 d\tau \,  [u\times \hat{B}(\hat{x} + \tau u)]_i
		\right)
		\exp(\ioverhbar u\cdot \hat{\pi}) \exp(\ioverhbar v\cdot \hat{x}) E_j \notag \\
        & = -v_i \hat{T}(u,v) E_j
		+ q\int_0^1 d\tau [u\times \hat{B}(\hat{x} + \tau u)]_i \hT(u,v) E_j \notag \\
        & = \hat{O}^A_{ij} + \hat{O}^C_{ij} , 
		\label{eq:pi-T-commutator}
	\end{align}
also using (4.25) from Ref.~\cite{serimaa}.
We can now look at the matrix elements of $ \mathcal{F} \left[ \hat{O}^A_{ij} \right]$, but it is of the form in~\cref{matels:v_i-term} except we have only one $\hat{E}$ operator, $\hat{E}(\hat{x})$.
Therefore,
    \begin{equation}
        \mathcal{F} \left[ \langle x' | \hat{O}^A_{ij} |x''\rangle \right]
        = \langle x' | \frac{\hbar}{i} \partial_{x_i} \left[
                E_j(x+ i \hbar \partial_p/2) \hat{W}
        \right] | x'' \rangle
        = \langle x' |  \frac{\hbar}{i}\left[
            E_{j,i}(x + i\hbar\partial_p/2) \hat{W}
            + E_j(x + i\hbar\partial_p/2) \partial_{x_i} \hat{W}
        \right] | x'' \rangle.
    \end{equation}
We see that the imaginary part of $E_{i,j}$ term cancels the imaginary part of the corresponding term in~\cref{matels:OD}, since when dividing by $i$, even powers of $i\hbar \partial_p$ should be kept.
For the second term, we can integrate by parts to show that
\begin{equation}
    \int_{-1/2}^{1/2} E_j(x + \tau u) \, d\tau
    + u_k \int_{-1/2}^{1/2} \tau E_{j,k}(x + \tau u) \, d\tau
    \\
    = \frac{1}{2} \big(E_j(x + u/2) + E_j(x - u/2)\big)
    \label{eq:ipb}
\end{equation}
Again keeping even powers of $i\hbar \partial_p$ since we are dividing by $i$,
\begin{equation}
    \operatorname{Im} \frac{\hbar}{i} E_j(x + i\hbar\partial_p) =
    -\frac{\hbar}{2} \big( E_j(x + i\hbar\partial_p/2) + E_j(x - i\hbar\partial_p/2)  \big)
    = - \hbar \tilde{E}_j - i\hbar^2 \partial_{p_k} \int_{-1/2}^{1/2} \tau E_{j,k}(x + i\hbar \tau \partial_p) \, d\tau.
    \label{eq:delta-E-tilde}
\end{equation}
This is the ``hidden momentum'' velocity correction term $\propto \bE\times \bs$ \cite{Babsonetal2009} (to be discussed more below), but with $\bE = \tilde{\bE} + \Delta \tilde{\bE}$ where $\Delta \tilde{\bE}$ is higher order in $\hbar$.

\end{widetext}
Following steps like those above, and using~\cref{eq:ipb} again, one readily expresses the matrix elements of $\hat{O}^C_{ij}$ in terms of those of $\hat{W}$.
The result is that
\begin{equation}
    \langle x' | \hat{O}^C_{ij} | x''\rangle
    =
    \epsilon_{inm} \tilde{B}_m (
        \tilde{E}_j
        + \Delta \tilde{E}_j
    ) \partial_{p_n}
    \langle x' | \hat{W} | x'' \rangle
\end{equation}
which, in the kinetic equation, will give
\begin{multline}
    \epsilon_{ijk} 	\epsilon_{inm} \tilde{B_m} (
    \tilde{E}_j
    + \Delta \tilde{E}_j
    ) \partial_{p_n} \sigma_k \hat{W} \\
    \propto
    \big[
        \big(
            (\tilde{\bE} + \Delta\tilde{\bE}) \times \bsigma
        \big) \times \tilde{\bB}
    \big] \cdot \partial_p \hat{W},
\end{multline}
i.e., it is the correction to the magnetic force due to the relation between momentum and velocity.
Here $\tilde{\bB}$ is defined in the same way as $\tilde{\bE}$, \cref{eq:E-tilde-def}.

\subsection{The spin transform}
\label{ssub:sub3}
It remains to evaluate the terms in the kinetic equation arising from the spin transform, proportional to
\[
	\epsilon_{ijk} s_i \sigma_k \big( (\hat{\bpi} \times \hat{\bE} - \hat{\bE} \times \hat{\bpi})_j\hat{W} + \operatorname{H.c.} \big)
\]
where $\operatorname{H.c.}$ denotes the Hermitian conjugate.
Using \cref{eq:pi-exp-pi} again we establish that
\begin{align}
	\hat{\pi}_k \hat{E}_l(\hat{x}) \hat{T} &= (\mathcal D_k\hat{T}) E_l(\hat{x} - u) \\
	\hat{E}_l(\hat{x}) \hat{\pi}_k \hat{T} &= \hat{E}_l(\hat{x}) \mathcal D_k\hat{T}
\end{align}
where $\mathcal D_k$ is the operator
\begin{equation}
	\mathcal D_k =
        \frac{\hbar}{i} \frac{\partial }{\partial u_k}
        + \frac{v_k}{2}
        - q \int_0^1 d\tau \, (1- \tau) [u\times \hat{B}(\hat x + u\tau)]_k.
\end{equation}

Hence the product $(\hat{\bpi} \times \hat{\bE} - \hat{\bE} \times \hat{\bpi})_j \hat{T}$ is given by
\begin{widetext}
    \begin{equation}
        \epsilon_{jkl}(\hpi_k \hE_l - \hE_j \hpi_k) \hat{T}
		= \epsilon_{jkl}(\hpi_k \hE_l + \hE_k \hpi_l) \hat{T}
        = \epsilon_{jkl} \left[
            \frac{\hbar}{i}  \left(  \pd{ \hat{T} }{u_k}  \hat{E}_l(\hat{x} - u)
                + \hat{E}_l(\hat{x}) \pd{\hat{T}}{u_k}
            \right)
            + v_k \hat{E}_l (\hat{x})\hat{T}
        - 2 I_k  E_l(\hat{x})\hat{T}  \right] , 
        \label{eq:pi-E-T}
    \end{equation}
\end{widetext}
where 
\begin{equation}
	I_k = \int_0^1 d\tau \, (1- \tau) [u\times \hat{B}(\hat x + u\tau)]_k.
\end{equation}
Differentiating $\hat{E}_l(\hx) \hT = \hT \hat{E}_l(\hx - u)$ with respect to $u_k$ we obtain
\begin{equation}
	\hat{E}_l(\hat{x}) \pd{}{u_k} \hat{T} = \pd{\hat{T}}{u_k} \hat{E}_l(\hat{x} - u) - \hat{T} \hat{E}_{l,k} (\hat{x} - u).
\end{equation}
so that
\begin{multline}
	\pd{ \hat{T} }{u_k}  \hat{E}_l(\hat{x} - u) + \hat{E}_l(\hat{x}) \pd{\hat{T}}{u_k}
	= 2\hat{E}_l(\hat{x}) \pd{\hat{T}}{u_k} + \hat{T} \hat{E}_{l,k}(\hat{x} - u) \\
    = 2\hat{E}_l(\hat{x}) \pd{\hat{T}}{u_k} + \hat{E}_{l,k}(\hat{x}) \hat{T}.
    \label{eq:E-dTdu}
\end{multline}
Putting this into \cref{eq:pi-E-T}, all terms are similar to ones already seen, and we will provide only an outline of the remaining calculations.
The operator of the type $\hat{E}_l(\hat{x}) \pd{\hat{T}}{u_k}$ we can handle as in~\cref{eq:dT-du-integral}, resulting in 
\[
	\sim p_k \tilde{E}_j (x-i\hbar \partial_p/2) W , 
\]
and following the argument leading to~\cref{eq:delta-E-tilde}, when taking the real part, we will get 
\begin{equation}
	\sim p_k (\tilde{E}_l + \Delta \tilde{E}_l) W.
\end{equation}
For the term proportional to $v_k$ in \cref{eq:pi-E-T}, we use that $\mathcal F : v_k \mapsto i\hbar \partial_{x_k} \delta (x)$ as leading up to \cref{matels:v_i-term}.
When taking the real part, the term where the derivative acts on $E$ cancels with the derivative term from \cref{eq:E-dTdu} and what remains is
\begin{equation}
	\sim i \hbar \left( E_l(x + i\hbar\partial_p/2 ) - E_l(x - i\hbar\partial_p /2 ) \right) \partial_{x_k} W
\end{equation}
which can be simplified using \cref{eq:E-imag-part}.

Following steps like those for $\hat{O}^B$ (\cref{eq:O-B} and following), the final term in \cref{eq:pi-E-T}, proportional to $I_k$, will give us a contribution of the form
\begin{widetext}
\begin{equation}
	-\int_{-1/2}^{1/2} d\tau \, \left(\frac{1}{2}-\tau \right) [u\times B(x + \tau u)]_k 
	E_l(x - u/2) \hat{T}.
\end{equation}
When taking the real part of the Fourier transform of this, there will be two terms,
    \begin{equation}
        \sim  \int_{-1/2}^{1/2} \tau [ i\hbar \partial _p\times B(x + i\hbar \tau \partial_p )]_k \, d\tau \
        \left ( E_l(x + i \hbar \partial_p/2 ) + E_l(x - i \hbar \partial_p/2  ) \right) W
        \propto
        \Delta \tilde{\bp} \times (\tilde{\bE} + \Delta\tilde{\bE}) W
    \end{equation}
    which is in line with previous generalisations, and
    \begin{equation}
        \sim \int_{-1/2}^{1/2}
        [ i\hbar \partial_p\times B(x + i\hbar \tau \partial_p )]_k \, d\tau \,
        \left ( E_l(x - i \hbar \partial_p/2 ) - E_l(x + i \hbar \partial_p/2  ) \right) W
        \propto
        (\tilde{\bB} \times \hbar \partial_p ) \times (\tilde{\bE} \hbar \xpleftright ) W
    \end{equation}
which is of a new type.
However, both are order $\hbar^2$, so their limits could not have been found in previous works that were either $\hbar$ or didn't include the spin-orbit interaction.

Thus, the spin torque is modified to add a term proportional to
$(\tilde{\bp} + \Delta\tilde{\bp}  ) \times (\tilde{\bE} + \Delta \tilde{\bE}) + \frac{q}{2}(\tilde{\bB} \times \hbar \partial_p ) \times \cdot (\tilde{\bE} \hbar \xpleftright)$.
The overall sign and coefficient of this term is found by matching its long-scale limit to the model in Ref.~\cite{asenjo2012semi}

\subsection{Conclusion}

\newcommand{\xpleftrightn}{\ensuremath{\overset{\leftarrow}{\nabla_x} \cdot \overset{\rightarrow}{\nabla_p} }}
The evolution equation for $f$ follows directly by taking the expectation value of the evolution equation for $\hat{W}$.
It is
    \begin{multline}
        \partial_t f
        + \frac{1}{m} \left(\bp + \Delta\tilde{\bp} + \frac{\mu_e}{2mc^2} (\tilde{\bE} + \Delta\tilde{\bE}) \times (\bs + \nabla_s) \right) \cdot \nabla_x f \\
        + q \left[ (\tilde{\bE}
            + \left(
                \bp + \Delta \tilde{\bp}
                + \frac{\mu_e}{2mc^2} (\tilde{\bE} + \Delta\tilde{\bE}) \times (\bs + \nabla_s)
            \right) \times \tilde{\bB}
        \right ] \cdot \nabla_p f
        + \mu_e \left(
            \tilde{\bB} - \frac{\bp + \Delta\tilde{\bp}}{2mc^2} \times \tilde{\bE}
            \right) \cdot (\bs + \nabla_s)
            (\xpleftrightn) f \\
            + \frac{2\mu_e}{\hbar} \left[\bs \times \left(
                    \tilde{\bB} + \Delta\tilde{\bB}
                    - \frac{1}{2mc^2} \left(\tilde{\bp} + \Delta\tilde{\bp}  \right) \times (\tilde{\bE} + \Delta \tilde{\bE})
                    - \frac{q}{4mc^2}(\tilde{\bB} \times \hbar \partial_p ) \times \cdot (\tilde{\bE} \hbar \xpleftrightn)
            \right) \right] \cdot \nabla_s f
        = 0.
        \label{eq:full-evol}
    \end{multline}
For reference, the notation used is
\begin{align}
    \tilde{\bE}(\bx) &
        = \int_{-1/2}^{1/2} \bE(\bx + i\hbar \tau \nabla_p /2) \, d\tau
        = \Big[\bE \int_{-1/2}^{1/2} \cos( \hbar\tau\xpleftrightn ) \, d\tau \Big](\bx)
        \\
    \Delta \tilde{\bE}(\bx) &
        = i\hbar \int_{-1/2}^{1/2} \tau \bE(\bx + i\hbar\tau \nabla_p) \, d\tau (\xpleftrightn)
        = -\hbar \Big[ \bE \int_{-1/2}^{1/2} \sin ( \hbar \tau \xpleftrightn) \, d\tau \, \xpleftrightn\Big](\bx)
        \\
    \Delta \tilde{\bp} &
        = -iq\hbar \int_{-1/2}^{1/2} \tau \bB(\bx + i\hbar \tau \nabla_p) \, d\tau \times \nabla_p
        = q\hbar \Big[\bB \int_{-1/2}^{1/2} \sin(\hbar\tau \xpleftrightn) \, d\tau \Big](\bx) \times \nabla_p
\end{align}
with $\tilde{\bB}$, $\Delta\tilde{\bB}$ defined in the same way as $\tilde{\bE}$, $\Delta\tilde{\bE}$.
Functions of operators are defined in terms of their Taylor series.
\end{widetext}

There are five new terms in \cref{eq:full-evol} as compared to Ref.~\cite{ZamanianNJP}.
In order, they represent: the ``hidden momentum'' (see below) correction to the velocity in the diffusion and magnetic force terms, respectively;
the spin-orbit force;
the spin torque due to spin-orbit interaction, including the Thomas precession;
and the last term, which is non-linear in the fields and higher-order in $\hbar$.
This term lacks an analog in Ref.~\cite{asenjo2012semi}; the others are short-scale generalizations of terms found in Ref~\cite{asenjo2012semi} similar to how Ref.~\cite{serimaa} generalizes the Vlasov equation.

We have denoted the electron magnetic moment by $\mu_e$ to allow for an anomalous magnetic moment, i.e., $\mu_e = \frac{g}{2} \mu_B$ where $\mu_B$ is the Bohr magneton and $g = 2 + \frac{\alpha}{\pi} + \ldots$.
That $g \neq 2$ leads to a mismatch between the cyclotron frequency and Larmor frequency, resulting in non-classical resonances~\cite{PhysRevLett.101.245002}, as will be seen below.
\label{sec:calculations-conc}

The system is closed with Maxwell's equations with polarization and magnetization,
\begin{align}
    \nabla \cdot \bE  &
        = \frac{1}{\epsilon_0}( \rho_f - \nabla \cdot \mathbf P ) \\
    \nabla \times \bB &
        = \mu_0 \mathbf{J}_f
        + \mu_0 \nabla \times \mathbf{M}
        + \frac{1}{c^2} \frac{\partial \bE}{\partial t}
        + \mu_0 \frac{\partial \mathbf P}{\partial t}.
\end{align}
The free charge $\rho_f$ is given by
\begin{equation}
    \rho_f = q \int f \, d\Omega,
\end{equation}
introducing $d\Omega = d^3p \, d^2 s$.
Finding the continuity equation for $\rho_f$ using \cref{eq:full-evol}, the free current $\mathbf{J}_f$ is given by
\begin{equation}
    \mathbf{j}_f = q\int \mathbf{v} f \, d\Omega = q\int \left( \frac{\bp}{m} + \frac{3\mu_e \bE \times \bs}{2m^2c^2} \right) f \, d\Omega
\end{equation}
and while it may appear strange that we do not have $\bp \parallel \mathbf{v}$, this is the function on phase space in Weyl correspondence with $\hat{\mathbf{v}} = \frac{i}{\hbar} [\hat{H}, \hat{\bx} ]$.
This is an example of ``hidden momentum'', common to systems with magnetic moments~\cite{PhysRevLett.18.876,PhysRev.171.1370,PhysRevLett.20.343,Babsonetal2009}, and is found already in the long-scale length limit~\cite{asenjo2012semi}, and also in Ref.~\cite{PhysRevE.96.023207}.

Finally, the polarization and magnetization densities are given by
{\allowdisplaybreaks
\begin{align}
    \mathbf P & = -3\mu_e \int \frac{\bs \times \bp}{2mc^2} f \, d\Omega \\
    \mathbf M & =  3\mu_e \int \mathbf s f \, d\Omega.
    \label{eq:magnetization}
\end{align}
}
Here, a factor $3$ appears due to the normalization of the spin transform; the magnetization is proportional to the expectation value of $\bsigma$.

It should be noted that a model very similar to \cref{eq:full-evol} was recently presented by Hurst \emph{et al.}~\cite{hurst2017phase}, starting from the same Hamiltonian.
Their model, however, is formulated in terms of a scalar quantity $f_0$ and a vector quantity $\mathbf{f}$, the mass and spin densities, respectively.
Mathematically, this is writing the Hermitian matrix-valued Wigner function in the basis $\{ I, \bsigma \}$, and the correspondence between models is that $f_0 = \int f \, d^2s$ and $\mathbf{f} = 3\int \bs f \, d^2s$.
Hurst \emph{et al.} find the same charge and current densities as we do%
~\footnote{Taking into account that their convention for the sign of the charge is different from ours.}
 including the same ``hidden momentum'' in the free current or velocity.
 They also find the new non-linear term in the spin torque.

 Hurst \emph{et al.}~\cite{hurst2017phase} derive their model using a completely different method, based on a gauge invariant expression for the Moyal bracket~\cite{mueller1999product}.
That our results agree, up to trivial transformations, lends confidence to the results.  We return to this in \cref{sec:discussion}.

\section{Linear waves}
\label{sec:dispersion}

To illustrate the usefulness of \cref{eq:full-evol} we study linear waves in a homogeneous plasma.
We consider two cases: first electrostatic waves in a magnetized plasma, then electromagnetic waves in an unmagnetized plasma.
From now on, we will let $c = 1$ to simplify the notation.

The momentum is expressed in cylindrical coordinates $\bp = (p_{\perp},\varphi_p,p_z)$ and the spin in spherical coordinates $\mathbf{s} = \sin{\theta_s}\cos{\varphi_s}\hat{\mathbf{z}}  + \sin{\theta_s}\sin{\varphi_s}\hat{\mathbf{y}} +   \cos{\theta_s}\hat{\mathbf{z} }$.
In the linearization of \cref{eq:full-evol} variables are separated according to $f = f_0(p_\perp, p_z, \theta_s) + f_1(x,p,s,t)$, $\bE = \bE_1$ and $\bB = B_0 \hat{\textbf{z}} + \bB_1$, where the subscripts $0$ and $1$ denote unperturbed and perturbed quantities respectively.
\cref{eq:full-evol} then becomes
\begin{widetext}
\begin{multline}
    \label{eq:linearized}
    \frac{\partial f_1}{\partial t} + \frac{\bp}{m} \cdot \nabla_x f_1+ \Big(\frac{\bp}{m}\times \bB_0\Big) \cdot \nabla_pf_1 + \frac{2\mu_e}{h}(\bs\times \bB_0)\cdot \nabla_s f_1 \\
    = -q\left[
        \tE + \left( \frac{\bp + \Delta \tilde{\bp}}{m}
            + \frac{\mu_e}{m}(\tE+ \Delta\tE)\times (\bs+ \nabla_s)
        \right)\times \mathbf{B}_0  + \frac{\bp }{m}\times \tB_1
    \right]\cdot \nabla_pf_0\\
    +\mu_e\nabla_x \Bigg[ \Big(\tB_1 -\frac{\bp \times \tE}{2m}
        \Big)\cdot (\bs + \nabla_s)
    \Bigg]\cdot \nabla_pf_0 
    \\
    - \frac{2\mu_e}{\hbar} \bs\times \Bigg[\tB +\Delta \tB
        -\frac{q\hbar^2}{4m}
        \big( \mathbf{B}_0 \times \nabla_p
            \big)\times \big(\tE   \ensuremath{\overset{\leftarrow}{\nabla_x} \cdot \overset{\rightarrow}{\nabla_p} }
        \big)
    - \frac{(\bp + \Delta \Tilde{\bp}) \times (\tE+ \Delta \tE) }{2m}\Bigg]
    \cdot \nabla_sf_0.
\end{multline}

In general the background distribution can be divided into its spin-up and spin-down components~\cite{lundin2010linearized}
\begin{equation}
    f_0(p_\perp, p_z, \theta_s)
    = \frac{1}{4\pi} \sum_{\nu = \pm 1} F_{0\nu}(p_\perp, p_z) (1 + \nu \cos \theta_s).
\end{equation}
In the sums below it will be implicit that $\nu$ takes the values $\pm 1$.

 \subsection{Electrostatic waves}

In this geometry we consider: $\mathbf{B}_0 = B_0\hat{\mathbf{z}}$, $\mathbf{k} = k\hat{\mathbf{z}}$ and $\mathbf{E} = E\hat{\mathbf{z}}$.
The perturbed parameters follow the plane-wave ansatz according to $f_1= \Tilde{f}e^{i\textbf{k}\cdot \textbf{x} -\omega t}$.
The left hand side of \cref{eq:linearized} can then be written
\begin{equation} \label{LHS}
    \Big[
   i\omega -i\frac{kp_z}{m}+ \omega_{ce}\frac{\partial}{\partial \varphi_p}+ \omega_{cg}\frac{\partial }{\partial \varphi_s}
    \Big]f_1=RHS,
\end{equation}
where $\omega_{ce}$ is the cyclotron frequency, $\omega_{cg} = \frac{g}{2}\omega_{ce}$ and $RHS$ is the right hand side of \cref{eq:linearized}.

Making an expansion of $\tilde{f}_1$ in eigenfunctions of the operators of the right hand side of \cref{eq:linearized} \cite{lundin2010linearized} we write
\begin{equation}
    \label{Eigen function}
    \tilde{f}_1(\bp, \bs) = \sum_{\alpha, \beta}\frac{1}{2\pi}g_{\alpha, \beta}(p_{\perp}, p_z, \theta_s) e^{-i(\alpha\varphi_s+\beta \varphi_p)}.
\end{equation}
Substituting \cref{Eigen function} in \cref{LHS}, we can now express $f_1$ in terms of $f_0$
\begin{equation}
    f_1 = \frac{A}{\omega - kp_z/m}
    + \frac{B_{+} e^{i(\varphi_s - \varphi_p)} }{\omega - kp_z/m + \Delta\omega_{ce}}
    + \frac{B_{-} e^{-i(\varphi_s - \varphi_p)}}{\omega - kp_z/m - \Delta\omega_{ce}},
\end{equation}
where
    \begin{align}
        A & = \sum_\nu -iq\tilde{E}\frac{\partial F_{0\nu}}{\partial p_z}\\
		B_{\pm} & = i\mu_e \sum_\nu \sin{\theta_s} \left( -\frac{q B_0}{4m}(\tilde{E} + \Delta \tilde{E}) \frac{\partial F_{0\nu}}{\partial p_{\perp}}
           - \frac{\nu p_{\perp}}{2m\hbar}(\tilde{E} + \Delta \tilde{E}) F_{0\nu}
           \pm \frac{ kp_{\perp}\tilde{E}}{4m} \frac{\partial F_{0\nu}}{\partial p_z}
	   \mp \frac{ \nu k \hbar qB_0 \tilde{E}}{4m} \frac{\partial^2 F_{0\nu}}{\partial p_{\perp}p_z } \right)
    \end{align}
and $\Delta\omega_{ce}=\omega_{cg}-\omega_{ce}$.
Next, noting that the magnetization current $\mathbf{J}_m = \nabla \times \mathbf{M} = 0$, we calculate the total current $\textbf{J}=\textbf{J}_f + \textbf{J}_p$, where $\textbf{J}_f$ and $\textbf{J}_p$ are the free and polarization current respectively,  in order to obtain the dispersion relation.
Note that the operators contained in $\tE$ and $\Delta \tE$ act on $F_0$ as translation of the kinetic momentum, see \cref{app:translation} for more details.
Introducing the notation
\begin{subequations}
\begin{align*}
    p_q &= \frac{\hbar k}{2}\\
    \qquad
    \pres^\pm& = \frac{m \omega}{k} \pm p_q
    \end{align*}
\end{subequations} 
the dispersion relation is then
\begin{equation}
    \label{Dispersion relation}
    \epsilon(k, \omega) = 1 + \chi_{fL} + \chi_{pL} = 0,
\end{equation}
where
\begin{subequations}
    \begin{align}
        \chi_{fL} & =
            \frac{4\pi^2q^2}{p_q k^2 m} \int d^2p \,
            \left(
                \frac{1}{\pres^+ - p_z}
                - \frac{1}{\pres^- - p_z}
            \right)
            F_{0\nu}
            \\
        \chi_{pL} & = - \frac{\pi^2\mu_e^2 }{km^2} \sum_{\pm,\nu} \int d^2p \bigg[
            \frac{p_{\perp}^2}{m\hbar}(\nu \mp 1) - \omega_{ce} (1\mp 2\nu )
        \bigg]
        \bigg(
              \frac{1}{\pres^\mp - p_z + m\Delta \omega_{ce}/k }
            - \frac{1}{\pres^\pm - p_z - m\Delta \omega_{ce}/k }
        \bigg)F_{0\nu}.
    \end{align}
    \label{eq:chifL}
\end{subequations}
\end{widetext}
Here $\chi_{fL}$ and $\chi_{pL}$ are the contributions due to the free and polarization currents, respectively.
We stress that in the first factor in $\chi_{pL}$, the sum over $\pm$ and $\nu$ corresponds to four terms in total.
Below we will use the indices $f$ and $p$ with the same meaning also for other quantities.
We note that $\chi_{fL}$ is the same susceptibility as for spinless quantum plasmas~\cite{eliasson2010dispersion}.
However, $\chi_{pL}$ generalizes previous results~\cite{asenjo2012semi} to cover also the short-scale physics.

Next, in order to examine how the spin affects the dynamic of the plasma, the Landau damping will be calculated and the contributions of $\chi_{fL}$ and $\chi_{pL}$ will be compared.
Starting from \cref{Dispersion relation}, we Taylor expand $\epsilon(k,\omega_r + i\gamma$) around $\omega_r$ \cite{nicholson1983introduction}, which leads to
\begin{equation}
    \label{Landau damping}
    \gamma = - \frac{\Im \epsilon(k, \omega_r)}{\partial (\Re \epsilon)/\partial \omega|_{\omega=\omega_r} }
    \quad \text{for} \quad
    |\gamma| \ll |\omega_r|.
\end{equation}

\begin{figure}
    \centering
    \includegraphics[width=\linewidth]{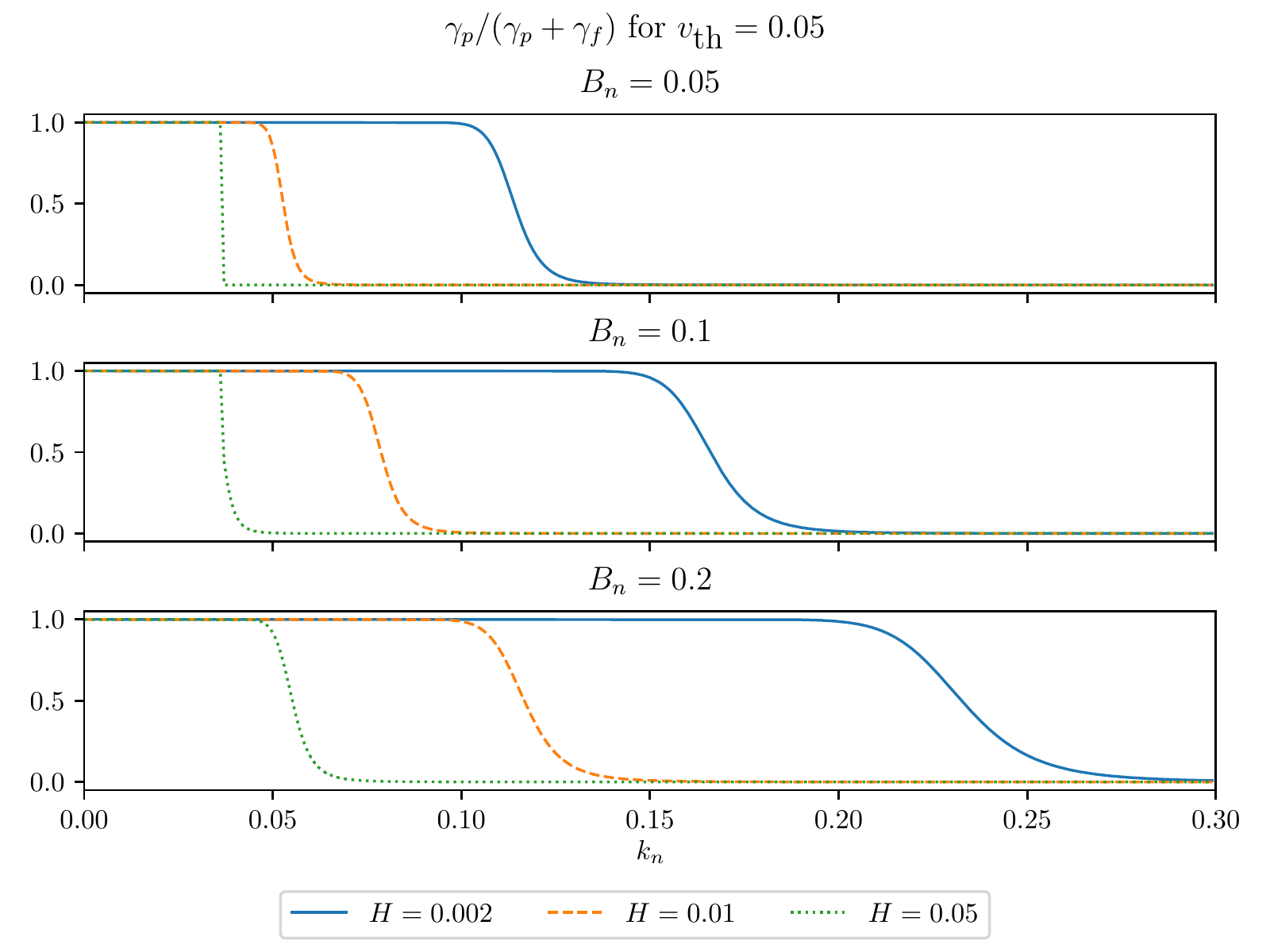}
    \caption{The fraction $\Gamma$ is plotted versus the normalized wave number $k_n=kv_{th}/\omega_p$ for different values of $B_n=\mu_BB_0/m$ and $H = \hbar\omega_p/m v_\textnormal{th}^2$ with thermal speed $v_{th}=0.05c$   }
    \label{fig:damping}
\end{figure}

\begin{widetext}
Analogously to \cref{Dispersion relation}, $\Im\epsilon$ can be divided into $\Im  \epsilon = \Im(\chi_{fL}) + \Im(\chi_{pL})$.
Using the Plemelj formula, we get
    \begin{subequations}
        \begin{align}
            \Im \chi_{fL } & = \frac{8\pi^3 q^2 m }{\hbar k^3} \int dp_{\perp}p_{\perp} \Big[
            F_{0\nu}(p_z = \pres^+) - F_{0\nu}(p_z = \pres^-)
            \Big]\\
            \Im \chi_{pL} & = \frac{\pi^3\mu_e^2 }{k} \sum_{\pm,\nu} \int dp_{\perp}p_{\bot} \Big[
                \frac{p_{\bot}^2}{m\hbar}(\nu \pm 1) -\omega_{ce}(1 \mp 2\nu )
                \Big]
                  \Big[
                      F_{0\nu}(p_z = \pres^\mp + m\Delta\omega_{ce}/k)
                    - F_{0\nu}(p_z = \pres^\pm - m\Delta\omega_{ce}/k)
                \Big].
        \end{align}
    \end{subequations}
\end{widetext}
To see how $\Im\chi_{pL}$ contributes to $\gamma$ compared to $\Im\chi_{fL}$, we define
\begin{equation}
    \Gamma
    = \frac{\gamma_{p}}{\gamma_{p} + \gamma_{f}}
    = \frac{\Im \chi_{pL}}{\Im (\chi_{fL} + \chi_{pL})}.
\end{equation}
The next step is to specify $F_{0\nu}$.
For simplicity, we will consider a Maxwell-Boltzmann distribution $F_{0\nu} = C_\nu e^{-p^2/\pth^2}e^{-\nu \mu_e B_0/k_{B}T}$ where $C_\nu$ is the normalization constant according to $\int d^3p d^2 s \, F_{0\nu} = N_{0\nu}$, where $ N_{0\nu} $ are the densities of spin-up ($ \nu=+1 $) and spin-down particles ($ \nu=-1 $).
This condition implies the textbook result that the magnetization is proportional to $\tanh \frac{\mu_e B}{k_B T}$.
We then have
\begin{widetext}
    \begin{subequations}
        \label{Imaginary Susceptibility}
        \begin{align}
        \Im \chi_{fL} & = \frac{\sqrt{\pi} m^2\omega_p^2}{\hbar k^3\pth}
            \left[
                e^{-(\pres^-/\pth)^2}
                - e^{-(\pres^+/\pth)^2}
            \right]\\
            \Im \chi_{pL} & = \frac{\hbar ^2 \omega_p^2\sqrt{\pi}}{32m^2k\pth}
            \sum_{\pm}
            \Bigg[
                \frac{-\pth^2}{m\hbar} \left(
                    \tanh \frac{\mu_BB_0}{k_BT} \pm 1
                \right)
                - \omega_{ce} \bigg(
                    1 \pm 2 \tanh \frac{\mu_BB_0}{k_BT}
                \bigg)
            \Bigg] \notag \\
                  & \times \Big[
                  e^{-\big(\pres^\mp + m\Delta\omega_{ce}/k \big)^2/\pth^2}
                - e^{-\big(\pres^\pm - m\Delta\omega_{ce}/k \big)^2/\pth^2}
            \Big].
        \end{align}
    \end{subequations}
Finally, the real part of the dispersion relation $\omega_r(k)$  has to be specified explicitly in order to obtain $\Gamma$.
Assuming that $\chi_{pL}$ in \cref{Dispersion relation} is small compared to $\chi_{fL}$
\footnote{This will be the case at least if $\hbar \omega_p / m \ll 1$.}
we neglect it for the purposes of computing $\omega_r$.
This gives us
    \begin{equation}
        \epsilon_r \approx 1- \frac{m\omega_p^2}{k\omega \sqrt{\pi}\pth}
        \int dp_z\left[
            \frac{p_z/\hbar k- 1/2}{p_{res}^{+}-p_z }
            -    \frac{p_z/\hbar k+ 1/2}{p_{res}^{-} -p_z }
        \right] e^{-p_z^2/\pth^2}.
    \end{equation}
\end{widetext}
Assuming $\omega \gg kv_{th}/m \pm \hbar k^2/2m $, we can make a Taylor expansion of the denominators up to first non-vanishing order.
The real part of the dispersion relation $\omega_r(k)$ is then
\begin{equation} \label{Real Dispersion relation }
    \omega_r(k)= \pm \sqrt{\omega_p^2 + k^2v_{th}^2 + \frac{\hbar ^2k^4}{4m^2}}.
\end{equation}
Using this expression of $\omega$ in \cref{Imaginary Susceptibility}, we can plot $\Gamma$, see \cref{fig:damping}.
In general the spin contribution to the damping dominates (i.e., $\Gamma \to 1$) for long wavelengths, whereas the free current contribution dominates (i.e., $\Gamma \to 0$) for shorter wavelengths.
The transition between these regimes is dependent on the normalized magnetic field $B_n = \mu_B B / m$ and the quantum parameter $H = \hbar \omega_p / m v_\textnormal{th}^2$.
We have performed the computation in \cref{fig:damping} for various other values of $v_\textnormal{th}$, however, this reveals that the location of the transition region, expressed in $k_n = kv_\textnormal{th}/\omega_p$, is almost independent of $v_\textnormal{th}$.

\subsection{Electromagnetic waves}

In this geometry we look at electromagnetic waves in unmagnetized plasmas with
$\textbf{k} = k\hat{\mathbf{z}}$, $\mathbf{E} = E\hat{\mathbf{x}}$ and $\mathbf{B} = B\hat{\mathbf{y}}$.
We will allow for a background distribution depending on both $p_\perp$ and $p_z$, but assume it to be spin-independent, as there is no background magnetic field.
Dividing $f$, $E$ and $B$ into perturbed and unperturbed quantities, \cref{eq:linearized} becomes
\begin{multline}
    \left[
        \frac{\partial}{\partial t} + \bp \cdot \nabla_x
    \right] f_1
    = -q \Big(\tE + \frac{\bp}{m}\times \bB \Big)\cdot \nabla_p f_0 \\
        - \mu_e\nabla_x\Big[
            \left(\bs + \nabla_s\right)\cdot \Big(\bB - \frac{\bp \times \tE}{2m}\Big)
        \Big] \cdot \nabla_pf_0.
\end{multline}
Using an expansion of $f_1$ like that in \cref{Eigen function} $f_1$ is computed as,
\begin{equation}
    f_1 = \frac{-1}{\omega - kp_z/m} \left[ A\cos{\varphi_p} - B\sin{\varphi_p} - C\sin{\varphi_s}
    \right],
\end{equation}
where
\begin{align}
    A & = iq\tilde{E} \frac{\partial f_0}{\partial p_{\perp}} + i\frac{q\tilde{B}}{m} \Big( p_{\bot} \frac{\partial f_0}{\partial p_{z}}
        - p_{z} \frac{\partial f_0}{\partial p_{\perp}}
    \Big)\\
    B & = -\frac{k\mu_Bp_{\perp} }{2m} \tilde{E} \cos{\theta_s}\frac{\partial f_0}{\partial p_z}
\end{align}
and
\begin{align}
    C & = k\mu_e\Big(
        \tilde{B} + \frac{p_z}{2m}\tilde{E}
        \Big)
        \sin{\theta_s}\frac{\partial f_0}{\partial p_z}.
\end{align}
Next we calculate the total current $\mathbf{J}_{T} = \mathbf{J}_{f} + \mathbf{J}_{p} + \mathbf{J}_{m}$, which together with Maxwell's equations gives us the dispersion relation
\begin{equation}\label{Dispersion relation Transverse}
    1 - \frac{k^2}{\omega^2} + \chi_{fT} + \chi_{pT} + \chi_{mT} = 0,
\end{equation}
where
\begin{widetext}
    \begin{subequations}\label{Dispersion relation 2}
        \begin{align}
            \chi_{fT}&=- \frac{\omega_p^2}{\omega^2}- \frac{4\pi^2 q^2k^2}{m^3 \omega^2}\int d^2p\, p_{\perp}^2 \frac{f_0}{(\omega^2-kp_z/m)^2 - \hbar^2k^4/4m^2}
            \label{eq:electromagnetic-chif} \\
            \chi_{mT}&= \frac{-8\pi^2k^2\mu_e^2}{m\omega^2}\int d^2p \frac{k^2- \omega^2/2}{(\omega-kp_z/m)^2 - \hbar^2k^4/4m^2}f_0\\
            \chi_{pT}&= \frac{4\pi^2k^2\mu_e^2}{m}
            \int d^2p \frac{f_0  }{(\omega-kp_z/m)^2 - \hbar^2k^4/4m^2} \Big[1+ \frac{p_{\perp}^2}{4m^2}- \frac{\hbar^2 k^2}{8}-\frac{\omega p_z}{km} +\frac{p_z^2}{2m^2}
            \Big].
        \label{eq:electromagnetic-chim}
        \end{align}
    \end{subequations}
\end{widetext}
Here the indices $f,m,p$ stand for the contributions to the susceptibility due to the free, magnetization, and polarization currents, respectively.

To shed some further light on the dispersion relation \cref{Dispersion relation Transverse}, we consider the limit of the vanishing background pressure. We then immediately obtain
\begin{multline}
    \label{Dispersion relation 3}
    \omega^2 =
     k^2 + \omega_p^2
    \bigg[1 + \frac{\hbar^2k^2}{4m^2}
   \frac{\big( \omega^2-k^2-\hbar^2k^4/16m^2 \big)}{\omega^2-\hbar^2k^4/4m^2}
    \bigg]
\end{multline}
The quantum contribution to \cref{Dispersion relation 3} comes from the magnetic dipole force and the spin-orbit interaction coupling the transverse motion with the longitudinal degrees of freedom.
This is illustrated by the free-particle dispersion relation in the denominator of the second term.
While \cref{Dispersion relation 3} highlights the physics in a simple manner we need to stress that the quantum modification of the classical dispersion relation \cref{Dispersion relation Transverse} is of most interest for high density plasmas, in which case the Fermi pressure is important even if the Fermi pressure is small.
Thus, for realistic applications, we typically must use the full expressions \crefrange{eq:electromagnetic-chif}{eq:electromagnetic-chim} for the susceptibilies, rather than \cref{Dispersion relation 3}.

\section{Discussion}
\label{sec:discussion}

The model derived here, \crefrange{eq:full-evol}{eq:magnetization}, is a generalization of the governing  Eqs. of Asenjo \emph{et al.}~\cite{asenjo2012semi} allowing for short macroscopic scale lengths, i.e. of the order of the de Broglie length.
Alternatively, we can view the model as a generalization of Zamanian \emph{et al.}~\cite{ZamanianNJP} into the weakly relativistic regime.
A kinetic theory based on the same Hamiltonian has previously been constructed by Hurst \emph{et al.}~\cite{hurst2017phase}.
Their approach leads to two coupled equations for a scalar $f_{0}$ and a vector $\mathbf f$, see \cref{sec:calculations-conc}.
While this is a notable difference, nevertheless the evolution equations of Hurst \emph{et al.}~\cite{hurst2017phase} generalize Hurst \emph{et al.}~\cite{hurst2014semiclassical} in the same way as our results generalize those of Zamanian \emph{et al.}~\cite{ZamanianNJP}.
However, as Hurst \emph{et al.}~\cite{hurst2014semiclassical} concluded, their model is equivalent to that of Zamanian \emph{et al.}~\cite{ZamanianNJP}, we can similarly conclude that the model here is equivalent to that of Hurst \emph{et al.}~\cite{hurst2017phase}, apart from the small but significant detail that we include the contribution from the anomalous magnetic moment.

To demonstrate the usefulness of our model we have calculated the linear dispersion relation for two simple cases:
Electrostatic waves propagating parallel to an external magnetic field and electromagnetic waves in an un-magnetized plasma.
The electrostatic dispersion relation possesses two roots. The usual Langmuir root and a root induced by the spin resonances with $\omega \simeq \Delta \omega _{c}$~\cite{asenjo2012semi}.
Here we have focused on the Langmuir root, showing that even when the real part of the frequency is almost unaffected by the spin dynamics, the wave-particle damping can still be significantly modified.

For the electromagnetic wave mode we note that all the new terms can be expressed in terms of the longitudinal susceptibility. 
Physically it means that the spin-terms couples the longitudinal and the transverse degrees of freedom. 
However, for scale lengths much longer than the Compton length ($k\ll mc/\hbar $) this coupling is relatively weak. 

A new feature of the present theory as compared to previous works is seen already for linear wave propagation, and comes from the denominators  $(\omega -kp_{z}/m\pm \Delta\omega _{c}\pm \hbar k^{2}/2m)^{-1}$ seen in \cref{eq:chifL}.
We note the importance of not letting $\mu_e=\mu_B$ here, as otherwise we would get $\Delta\omega _{c}=0$.
The physical origin of the resonances corresponding to $(\omega -kp_{z}/m\pm \Delta\omega _{c}\pm \hbar k^{2}/2m)=0$ comes from energy momentum conservation written in the form
\begin{align}
    \hbar \omega_{2} & = \hbar \omega _{1} \pm \hbar \Delta \omega_{c} \pm \hbar \omega
    \label{Energy-q} \\
    \hbar k_{2} & = \hbar k_{1}\pm \hbar k \label{momentum-q}
\end{align}
Here $(k_{1},\omega _{1})$ and $(k_{2},\omega _{2})$ are the wavenumbers and frequencies of free particles before and after the interaction with the wave field, fulfilling the free particle dispersion relation $\omega_{1,2}=\hbar k_{1,2}^{2}/2m$.
The generalization from previous results (e.g., Refs.~\cite{eliasson2010dispersion,brodin2017nonlinear}) comes from the term $\pm \hbar \Delta \omega_{c}$ in (\ref{Energy-q}), corresponding to resonant particles gaining an amount $\hbar \omega_{c}$ of perpendicular kinetic energy, at the same time as the magnetic dipole energy drops by $\hbar \omega_{cg}$, or vice versa.
For our particular case of electrostatic waves propagating parallel to $\mathbf{B}_{0}$, the kinetic energy and magnetic dipole energy always change in opposite directions, as reflected by the appearance of $\Delta \omega_{c}$ in (\ref{Energy-q}).
However, in a more general setting, the energy relation will be of the form
\begin{equation}
    \hbar \omega_{2}=\hbar \omega_{1}\pm \hbar n\omega_{c}\pm \hbar \omega_{cg}\pm \hbar \omega
    \label{Energy-q2}
\end{equation}
whereas \cref{momentum-q} will be unaffected. Here $n$ is an integer and all sign combinations for the last three terms in (\ref{Energy-q2}) are possible.

Finally we point out a problem of future interest; to compute the ponderomotive force starting from \crefrange{eq:full-evol}{eq:magnetization}.
Previous results~\cite{stefan2011ponderomotive} indicates that the spin-orbit interaction can give a substantial contribution.
Hence it would be of interest to see to what extent the short-scale physics of the present model modify the results of Ref.~\cite{stefan2011ponderomotive}.

\begin{acknowledgments}
All authors would like to acknowledge financial support by the Swedish Research Council, grant number 2016-03806.
G.~Brodin, H.~Al-Naseri, and J.~Zamanian would also like to acknowledge support by the Knut and Alice Wallenberg Foundation through the PLIONA project.
\end{acknowledgments}


\appendix

\begin{widetext}
\section{}

\label{app:translation}

In this section we show how $\tE$ acts on $f_{0}(p)$ as a translation of momentum.
The same applies to $\tB$ since it has a similar form.
We can rewrite $\tE \,\partial f_0/\partial p_z$ as

{\allowdisplaybreaks
\begin{align}
    \label{First relation}
    \tE \frac{\partial f_0}{\partial p_z}
	&= \mathbf E(x)\int^{1/2}_{1/2}  d\tau \cos{(i\hbar k\tau \partial_{p_z}) \frac{\partial f_0}{\partial p_z}}
	=- \frac{2i \mathbf{E} }{\hbar k } \left( \frac{\partial}{\partial p_z} \right)^{-1}  \sin({i p_q \partial_{p_z}})
    \frac{\partial f_0}{\partial p_z} \notag \\
	&= \frac{\mathbf{E}}{2p_q} \Big( f_0(p_z +p_q)- f_0(p_z- p_q)
    \Big),
\end{align}
}
where $p_q= \hbar k /2$. However, for $\Delta \tE f_0$, we can use integration by parts
\begin{equation}
    \label{Second relation}
	\Delta \tE f_0 =-\hbar \mathbf{E} \int ^{1/2}_{1/2} d\tau \tau \sin{(i\hbar \tau k \partial_{p_z})} ik \frac{\partial f_0}{\partial p_z}
    = \frac{\mathbf{E}}{2}\Big( f_0(p_z+p_q) + f_0(p_z-  p_q)
    \Big)  -\tE f_0.
\end{equation}
For the case of $\tilde{\mathbf{E}} \frac{\partial f_0}{\partial p_{\perp}}$,
\begin{equation}
		\tilde{\mathbf{E}} \frac{\partial f_0}{\partial p_{\perp}} 
		= \mathbf{E} \int^{1/2}_{1/2}d\tau \cos{(i\hbar k\tau \partial_{p_z})} 
		\frac{\partial f_0}{\partial p_{\bot}}
        =\frac{\mathbf{E}}{2}\int_{1/2}^{1/2}d\tau
        \bigg[\frac{\partial f_0}{\partial p_{\perp}} \Big(p_z+ \frac{\hbar k\tau}{m}\Big)
            +\frac{\partial f_0}{\partial p_{\perp}} \Big(p_z- \frac{\hbar k\tau}{m}\Big)
        \bigg].
\end{equation}
Transforming the coordinates $(p_z,\tau)$ to $(u_{\pm},\tau')$, where $u_{\pm}= p_z \pm \hbar k\tau $
\begin{equation}
    \int^{\infty}_{-\infty}dp_z \tilde{\mathbf{E}}\frac{\partial f_0}{\partial p_{\perp}} = \frac{\mathbf{E}}{2}\int^{\infty}_{-\infty}du_{+} \frac{\partial f_0}{\partial p_{\bot}}(u_{+})\\
    + \frac{\mathbf{E}}{2}\int^{\infty}_{-\infty}du_{-} \frac{\partial f_0}{\partial p_{\perp}}(u_{-})
    = \mathbf{E}\int^{\infty}_{\infty}dp_{z} \frac{\partial f_0}{\partial p_{\perp}}(p_{z})
\end{equation}
\end{widetext}

\bibliography{Short-scale}{}

\end{document}